\def\wh{wormhole }
\def\beq{\begin{equation}}
\def\eeq{\end{equation}}
\def\bea{\begin{eqnarray}}
\def\eea{\end{eqnarray}}

\def\th{_{_{\rm th}}}

\documentstyle[aps,prl,epsfig]{revtex}
\begin{document}                                       
\title{Evolving wormhole geometries}
\author{Luis A. Anchordoqui\thanks{E-mail: doqui@venus.fisica.unlp.edu.ar}, 
Diego F. Torres, Marta L. Trobo}

\address{Departamento de F\'{\i}sica, Universidad Nacional de La Plata\\
C.C. 67, 1900, La Plata, Argentina\\}

\author{\&}

\author{Santiago E. Perez Bergliaffa\thanks{Permanent address: Departamento
de F\'{\i}sica, Universidad Nacional de La Plata, C.C. 67, 1900, La Plata,
Argentina}}

\address{International Center for Theoretical Physics\\ POBox 586,
I-34014, Trieste, Italy} 
\maketitle
\begin{abstract}
We present here analytical solutions of General Relativity that 
describe evolving wormholes with a non-constant 
redshift function. We show that the matter that threads these wormholes 
is not necessarily exotic. Finally, we investigate some 
issues
concerning WEC violation and human traversability in these
time-dependent geometries.

\noindent {\it PACS number(s):} 04.20.-q, 04.20.Jb

\end{abstract}

\section{Introduction}

In the last two decades,  there has been a revival on the study of
classical wormhole solutions either in General Relativity (GR) or in 
alternative theories of gravitation.
A cornerstone in this revival has been the paper of Morris and Thorne
\cite{motho} in which, 
they derived the properties that a space-time 
must have in order to hold up such a geometry. 
Let us recall
that the most salient
feature of these space-times
is that an embedding of one of their space-like sections in euclidean space
displays 
two asymptotically flat regions joined by a 
throat.
There are several reasons that support the interest in these
solutions. 
One of them is   
the possibility of constructing time machines \cite{mothoyu,fmnekty,fn}
which violate Hawking's chronology protection conjecture \cite{haw}.
Another reason is related to the nature of the matter that generates this 
solution: it must be ``exotic'', {\em i.e.}
its energy density takes negative values in some reference systems.
This entails violations of the weak energy
condition (WEC) which, although not based in any strong physical
evidence \cite{VISSER}, is deep enough in the body of relativists'
belief so as to abandon it without serious reasons. 
 These violations are 
in fact encoded in the
evolution of the expansion scalar, governed by Raychaudhuri's equation 
\cite{RAY}. 
As WEC is violated for any static spherically symmetric \wh and
has recently been shown to be the case also for any static \wh in GR
\cite{vis}, the attemps to get around WEC violations have led two
different branches of research. Firstly, there have been an increasing number of
works in non-standard gravity theories.
Brans-Dicke gravity supports static wormholes both in vacuum \cite{AGNESE}
and with matter content that do not violate the WEC by itself 
\cite{TORRES_WH}. Also, there are analysis in several others alternative
gravitational theories, such as
$R+R^2$  \cite{93tesis}, Einstein-Gauss-Bonnet \cite{94tesis} and
Einstein-Cartan models \cite{92tesis}. 
In all cases, the violation of WEC is a necessary condition for static \wh to exist,
although these changes in the gravitational action allows one to have normal
matter while relegating the exoticity to non-standard fields.
This lead to explore the issue of WEC violation in non-static situations
and see which is the case for dynamic, time-dependent wormholes.
The first of this kind of analysis, due to Roman \cite{ROMAN}, 
was devoted to the case of 
inflating Lorentzian wormholes. 
These are wormholes of the Morris-Thorne type
embedded in an inflating background, with all  non-temporal
components of the metric tensor multiplied by a factor of the form
$e^{2\chi t}$, being $\chi$ related to the cosmological
constant. The aim of Roman was to show that a microscopical \wh could
be enlarged by inflationary processes. However, this construction also 
violate the WEC.
The other works, due to Kar and Kar and Sahdev
\cite{KAR,KAR2},  pointed at testing whether,
within
classical
GR,  
a class of non-static,  non-WEC-violating wormholes could exists. 
They used a metric
with a conformal time-dependent factor, whose spacelike sections 
are $R \times S^2$
with a \wh metric, like the ones analyzed in \cite{motho}. 
In this work, we shall extend
this last analysis by studying more general cases of evolving
wormholes, given by a generic Morris-Thorne  metric, 
in the presence of matter
described by a given stress-energy tensor. To proceed further
we shall make two ans\"atze. Firstly, we 
give a specification of the 
redshift function $\Phi$
of Morris-Thorne consistent,
for any hypersurface of constant $t$,
with asymptotic flatness
and a metric free of horizons, as was the one made for the static case
in \cite{karsah}. Then we shall show how to choose a particular
functional form for the trace of the stress-energy tensor  matter in
order to get
different solutions for the conformal factor.  At this point, we shall
derive an analytical solution which we use later to analyze the 
WEC  violation scenario and some human tranversability criteria.

\section{Geometrical and topological features}

We adopt the following diagonal metric for the space-time:

\begin{equation}
ds^2 = \Omega(t) \, \{ - e^{2\Phi(r)} dt^2 + e^{2\Lambda(r)} dr^2 + 
r^2 d\Omega_{2}^2 \}
\label{metric}
\end{equation}
with $\Omega(t)$ a conformal factor, finite and positive defined
throughout the domain of $t$, and $d\Omega_{2}^2$, the $S^2$ 
line element as usual. When $\Omega (t)$ is constant, this general metric
stands for the ones firstly analyzed in the work of Morris and Thorne 
\cite{motho}.
When $e^{2\Phi} =1$, it reduces to the particular case studied by Kar 
\cite{KAR}.
In the spirit of \cite{karsah},
we make the Ansatz $\Phi = -\alpha/r$, where $\alpha$ is a positive constant 
to be determined. This choice guarantees that the redshift function is 
finite everywhere, and consequently there is no event horizon.
For the stress-energy tensor we take that of an imperfect
fluid 
\footnote{In what follows, greek 
indices run from 0 to 3, parenthesis denotes symmetrization,
$\xi_{\mu}=\sqrt {-g_{00} }\delta_\mu^0$ 
represents time-like unit vectors and 
$\zeta_{\mu}=\sqrt {g_{11} }\delta_\mu^1 $ 
space-like unit vectors in the radial direction respectively.}\cite{MM}, 
\begin{equation}                                 
\label{tensor}
T_{\mu\nu} = (\rho + p) \,\xi_{\mu} \xi_{\nu} 
+ p\,g_{\mu\nu} + \Delta p 
\left[ 
\zeta_{\mu} \zeta_{\nu} - \frac{1}{3}\, (g_{\mu\nu} 
+ \xi_{\mu} 
\xi_{\nu}) \right] + 2 \,q \,\zeta_{(\mu} \xi_{\nu)} 
\end{equation}
where $\rho$ stands for the energy density, $p$ for 
the isotropic fluid pressure, 
and $q$ for the energy flux. The anisotropy pressure $\Delta p$ is the
difference between the local radial ($p_r$) and lateral
($p_{\perp}$) stresses.  
With the foregoing assumptions, the nonvanishing Einstein's equations
are 
\begin{equation}
 \frac{3 \,\dot{\Omega}^2 \,e^{2\alpha/r}}{4\,\Omega^3} + \frac{2 \,\Lambda' \,
e^{-2\Lambda}}{r \, \Omega} + \frac{1}{r^2 \, \Omega} - \frac{e^{-2 \Lambda
}}{r^2 \,\Omega} = 8 \pi \rho
\end{equation}
\begin{equation}
\frac{3\,\dot{\Omega}^2 \,e^{2\alpha/r}}{4\,\Omega^3} + \frac{ 2 \,\alpha
\, e^{-2\Lambda}}{r^3\, \Omega} - \frac{\ddot{\Omega} \, e^{2\alpha/r}}
{\Omega^2}
- \frac{ 1}{\Omega \, r^2} + \frac{ e^{-2\Lambda} }{\Omega \, r^2} = 
8 \pi p_r
\end{equation}
\begin{equation}
\frac{3 \, \dot{\Omega}^2 \, e^{2\alpha/r}}{4\, \Omega^3} 
- \frac{\ddot{\Omega}\, e^{2\alpha/r}}{\Omega^2}
- \frac{e^{-2\Lambda} \,\alpha}{r^3\, \Omega} - \frac{\Lambda'\,e^{-2 \Lambda}}
{r \, \Omega} + \frac{\alpha^2 \,e^{-2 \Lambda}}{ r^4 \,\Omega} - 
\frac{ \alpha\, \Lambda'\,
e^{-2 \Lambda}}{\Omega \, r^2} = 8 \pi p_{\perp}
\end{equation}
\begin{equation}
\frac{ \dot{\Omega} \,\alpha }{r^2
\, \Omega^2}
= 8 \pi q \,\, e^{\Lambda} \, e^{ \alpha/r}
\label{flux}
\end{equation}
In the expressions given above, dashes denote derivatives with respect
to $r$ while dots, derivatives 
with respect to $t$. It seems convenient to assume
that the trace of the stress energy tensor will be a separated 
function of $t$ and $r$, 
\bea
T_{\mu}^{\mu} =
 \left[ -3 \frac{\ddot{\Omega}}{\Omega} +  \frac{3}{2}
\left(\frac{\dot{\Omega}}{\Omega}\right)^2 \right] 
\frac{e^{2\alpha/r } }{\Omega} +
\frac{1}{\Omega} \left\{ 
- 2 \, e^{-2\Lambda}\left[ \Lambda' \left(
\frac{2}{r} + \frac{\alpha}{r^2} \right) \right. \right.& + & \left. \left. 
\frac{e^{2\Lambda}}{r^2} -
\frac{1}{r^2} - \frac{\alpha^2}{r^4} \right] \right\} \nonumber \\
 & := &  \Upsilon (t)  \frac{e^{2\alpha/r}}{\Omega(t)}
\label{traza}
\eea
After separating variables, equation (\ref{traza}) can be rewritten as,
\begin{equation}
\label{5a}
-3 \frac{\ddot{\Omega}}{\Omega} +  \frac{3}{2}
\left(\frac{\dot{\Omega}}{\Omega}\right)^2 - \Upsilon  = {\cal C}
\end{equation}
\begin{equation}
\label{5b}
 2 \,e^{-2\alpha/r}\, e^{-2\Lambda}\left[ \Lambda' \left(
\frac{2}{r} + \frac{\alpha}{r^2} \right) + \frac{e^{2\Lambda}}{r^2} -
\frac{1}{r^2} - \frac{\alpha^2}{r^4} \right] = {\cal C}
\label{radial}
\end{equation}
where ${\cal C}$ is an arbitrary constant which for simplicity, we shall take
equal to 0. 
Performing the change of variables $\Lambda = {\rm ln} z$, equation
(\ref{radial}) transforms into a Bernoulli equation, 
\begin{equation}
\frac{- z'}{z} \left(\frac{2}{r} + \frac{\alpha}{r^2} \right) - 
\frac{z^2}{r^2} + \left(\frac{1}{r^2} + \frac{\alpha^2}{r^4} \right) = 0 .
\end{equation}
The additional change $y=z^{-2}=\exp{[-2\Lambda]}$, allows for the
general solution to be written as
\begin{equation}
\label{solucion}
y=e^{-2\Lambda} = \frac{\phi^2 + 9 \phi^3 + 32
\phi^5 + (\,57/2 + e^{2/\phi}\,{\cal K}\,) \phi^4}{(2 \, \phi + 1)^5}.
\end{equation}
with $\phi=r / \alpha$, and ${\cal K}$ an integration constant to be 
determined. 
Note that, far from the \wh  throat,
\beq
\lim_{r\rightarrow \infty} e^{-2 \Lambda} =1,
\eeq
or equivalently,
\beq
\lim_{r\rightarrow \infty} \frac{dz}{dr}=0,
\eeq
where $z(r)$ is the embedding function \cite{motho}.
Moreover, for any hypersurface of constant $t$ the mouths of the wormhole  
need to connect two asymptotically  flat spacetimes, thus the geometry at 
the wormhole's throat is severely constrained. Indeed, 
the definition of  the throat (minimum wormhole radius), 
entails for a vertical
slope of the embedding surface,
\beq
\lim_{r\rightarrow r\th ^+} \frac {dz}{dr} =
\lim_{r\rightarrow r\th ^+} \pm \sqrt{e^{2\Lambda} -1} = \infty;
\label{Martikahermosa}
\eeq
besides, the solution must 
satisfy the 
{\it flaring out condition} \cite{motho}, which stated
mathematically reads,
\beq
-\frac {\Lambda^\prime e^{-2\Lambda}}{(1-e^{-2\Lambda})^2} > 0.
\label{Martikaestamashermosaaun}
\eeq
Equations (\ref{Martikahermosa}) and (\ref{Martikaestamashermosaaun}) 
will be satisfied if and only if
\begin{equation}
\lim_{\phi \rightarrow \phi\th^{^+}} e^{-2\Lambda} = 0^+, 
\label{fo}
\end{equation}
thus, in order to fix the constant ${\cal K}$, we 
we must select a value for the 
dimensionless radius of the throat 
($\phi\th  >0$) such that Eq. (\ref{fo}) 
be satisfied. 
Nevertheless, the absolute size of the throat also depends 
on $\alpha$. As an example let us impose $\phi\th =2$ which, 
using (\ref{solucion}), sets ${\cal K} = -97.25\,e^{-1}$ and $r\th = 2 \,\alpha$.
The aforementioned properties of $\Lambda$, together with the 
definition of $\Phi$ and $\Omega$, entail that 
the metric tensor describes, for $\Omega (t)=$ constant,
two asymptotically flat spacetimes joined by a 
throat. Thus, equation (\ref{solucion}) represents an analytical solution
for an evolving \wh geometry, which, due to the choice we made for the
trace, is independent of $\Omega(t)$. To obtain the complete behavior
of the metric, we have to define $\Upsilon (t)$ in order to solve for
$\Omega (t)$ in equation (\ref{5a}). 
We take instead Eq. (\ref{5a}) as the definition 
of $\Upsilon(t)$ in what follows.                  

\section{Avoiding WEC violations}        

Having 
at our disposal an explicit \wh solution, given in the form of $y(\phi)$,
we can now turn to the question of whether it entails WEC violations.   
Of course, we have to note that, in the general case, 
it will ultimately depend
on the specific choice of $\Omega(t)$, but we use here the fact that, 
due to the form in which
the solution for $y(\phi)$ was computed, it is independent of the
explicit form of the conformal factor. Thus, we are able to study different
functional forms of $\Omega(t)$ for the same $y(\phi)$.
The WEC asserts that for any
time-like vector $v^\mu$:
\beq
T_{\mu\nu} v^\nu v^\mu \ge 0.
\eeq
The physical significance of this condition is that the observed energy density,
as measured by any time-like observer, has to be positive. 
The connection of WEC violations with the existence of wormholes
geometries has been widely 
studied before \cite{mothoyu,VISSER,vis,KAR}. In the
case of evolving wormholes,
Kar \cite{KAR} has demonstrated that there exist \wh solutions
which do not violate the WEC for a specific choice of the redshift function.
His analysis is indeed general, in the sense that
even without having worked out any explicit \wh solution, he
could conclude that there exist some Lorentzian
evolving \wh geometries with the required matter not violating the WEC.
We must recall however that Kar dealt with the particular case $\Phi
= 0$.
We pursue here to refine his analysis, by means of the solution
for the metric component $g_{11}$ given in the previous section.
With the stress-energy tensor of equation (\ref{tensor}), WEC entails
\beq
\rho \ge 0 \;\;\;\;\;\;\;\rho\,+p_r+2\,q \ge 0
\;\;\;\;\;\;\;\rho+p_\perp  \ge 0 .
\eeq
Starting from the Einstein equations, the three previous inequalities
can be written as follows:
\beq
f_1(t)+g_1(r) \ge 0 \;\;\;\;\;\forall(t,r)
\eeq
\beq
f_2(t)+g_2(r) +2\,q(r,t)\ge 0 \;\;\;\;\;\forall(t,r)
\eeq
\beq
f_3(t)+g_3(r) \ge 0 \;\;\;\;\;\forall(t,r)
\eeq 
where the functions $f_i$ and $g_i$ are as in Table \ref{tabla1}.
It is a trivial exercise to write all radial functions in 
terms of the $\phi$-variable by means of the definition
$\phi=r/\alpha$.
After this is done,
it is seen that these functions display an 
explicit dependence on the $\alpha$
parameter and on $\phi \th$, through ${\cal K}$. The radius of the throat
will be fixed when $\alpha$ and $\phi \th$ are simultaneously known.
These radial functions are plotted in Figs. 1,2 and 3
for different values of $\alpha$ and for $\phi \th =2$.
Note that, at or near the throat, some of them are negative. This just
reflects
the fact that no static wormholes can  be built, within GR,
without using exotic matter.  Far enough from the throat,
all functions tend
to zero, being strong the dependence on $\alpha$ of the shape
of the curves. Also from the 
figures, we may note that if we demand that the time-dependent
functions $f_1,f_2$ and 
$f_3$
be greater than the modulus of the minimum of each of the
corresponding radial functions, all WEC inequalities will be fulfilled.
In the case of the second inequality, the term coming from the energy flux 
must be taken into account to determine the non-exotic region. 
In Table \ref{tabla2}, we provide several examples of possible solutions
for $\Omega(t)$. The procedure to obtain these particular functions
as solutions for the metric is, as stated, 
to define a suitable functional form for the temporal part of the 
trace of the stress-energy
tensor for matter, i.e. $\Upsilon (t)$. 
This values are also given together with the values of
the temporal functions $f_i$ for each choice. This entries
would allow comparison with WEC inequalities in order to search for
constraints on the values 
of the constants involved in each choice of $\Omega(t)$, 
to be fulfilled in
order to
obtain non-exotic matter  at the throat of the evolving
wormhole.
Finally, the domain of $t$ for which $\Omega$ is finite and without zeros
is also given in Table \ref{tabla2}. The comments made by Kar \cite{KAR}
when he considered different $\Omega (t)$ functions in his scheme
are totally
applicable here. For instance, we also have exponentially expanding or
contracting phases -for the first two solutions- and a \wh -like
universe (with a bang and a crunch) for $\Omega(t)=\sin^2(\omega t)$.
In our scheme, the particular choice of $\Omega(t)$ will determine
the energy flux, through the field equation (\ref{flux}). 
The sign of the energy flux (which depends on $\dot\Omega$) will
decide in turn if the \wh is 
attractive or repulsive, in the sense explained in
\cite{ROMAN}. For large enough and
positive $q$, the second WEC inequality  will
be fulfilled easily. Morover, 
this would imply a quick process of expansion
which in turn favors a possible trip by diminishing the tidal forces, as
we shall show in the next section.

\section{Traversability criteria}

We shall deal 
now with the possibility that an explorer might enter into 
the kind of evolving \wh
studied here, pass through the tunnel, 
and exit into an external space again 
without being crushed during the journey. We are not going to 
discuss here the stability of the wormhole, being this
beyond the scope of this paper. Instead, we              
content ourselves with an analysis of the tidal gravitational forces 
that an infalling radial observer must bear during the trip. The extension
to a 
non-radial motion follows from the recipe given in 
\cite{VisserBOOK} (Chap.
13).
We recall the desired 
traversability criteria concerning the tidal forces that
an observer should feel during the journey: they must not exceed the
ones 
due to Earth gravity \cite{GRAVITATION}. From now on, 
the algebra will be simplified by switching to an 
orthonormal reference frame in which
$g_{\hat\mu \hat \nu}=\eta_{\hat \mu \hat \nu}$.
In terms of this
basis, the proper reference frame is given by,
\begin{equation}
e_{\hat{0}'} = \gamma \,e_{\hat{t}} \mp \gamma \,\beta\, e_{\hat{r}}
\hspace{1cm} e_{\hat{1}'} = \mp \,\gamma \,e_{\hat{r}} + \gamma \,\beta\,
e_{\hat{t}} \hspace{1cm}
e_{\hat{2}'} = e_{\hat{\theta}} \hspace{1cm} e_{\hat{3}'} = e_{\hat{\phi}} 
\end{equation}
where $\gamma$ and $\beta$ are the usual parameters of the Lorentz
transformation formula. Thus, stated mathematically the 
traversability criteria read,
\begin{equation}
|R_{\hat{1}'\hat{0}'\hat{1}'\hat{0}'}| = \left| \frac{1}{\Omega}
\left\{ \frac{\alpha^2}{e^{2\Lambda}\,r^4} -
\frac{2\,\alpha}{e^{2\Lambda}r^3} -
\frac{\ddot{\Omega}\,e^{2\alpha/r}}{2\,\Omega} + \frac{\dot{\Omega}^2
e^{2\alpha/r}}{2\, \Omega^2} - \frac{\alpha\, \Lambda'}{e^{2\Lambda}
r^2} \right\} \right|
\leq  
\frac{g_{_{\oplus}}}{2m} \approx \frac{1}{(10^{10} {\rm cm})^2} 
\label{t1}
\end{equation}
\begin{displaymath}
|R_{\hat{2}'\hat{0}'\hat{2}'\hat{0}'}| = |R_{\hat{3}'\hat{0}'\hat{3}'\hat{0}'}| =
\gamma^2 |R_{\hat{\theta}\hat{t}\hat{\theta}
\hat{t}}| + \gamma^2 \beta^2 |R_{\hat{\theta}\hat{r}\hat{\theta}\hat{r}}| + 2 \gamma^2
\beta |R_{\hat{\theta}\hat{t}\hat{\theta}\hat{r}}| = 
\end{displaymath}
\begin{equation} 
 \left| \frac{\gamma^2}{\Omega} \left\{ \frac{\dot{\Omega}^2 \,
e^{2\alpha/r}}{2\,\Omega^2} -
\frac{\ddot{\Omega}\,e^{2\alpha/r}}{2\,\Omega} +
\frac{\alpha}{e^{2\Lambda}\,r^3} + \beta^2\, \left[
\frac{\dot{\Omega^2} \,e^{2\alpha/r}}{4 \,\Omega^2} +
\frac{\Lambda'}{e^{2\Lambda}\,r}\right] +\beta\,\frac{ \alpha\, \dot{\Omega}
e^{\alpha/r}}{r^2\,\Omega\,e^{\Lambda}}\right\}\right| 
\leq \frac{g_{_{\oplus}}}{2m}.
\label{t2}
\end{equation}
Note that, contrary to the static case, as the geometry is changing with 
time, the tidal forces that the hypothetical observer feels
also depend on time. 
The first of these inequalities 
could be interpreted as a constraint
on the gradient of the redshift, while the second acts as a constraint
on the
speed of the spacecraft. It is important to stress that we have here
a term associated with the energy flux, the last one of the left hand side
in equation (\ref{t2}).

Let us take as an example the case of an expanding wormhole, with 
$\Omega= e^{2 \omega t}$ and domain for $t$ in the interval 
$(-\infty, \infty)$. 
Now, equations
(\ref{t1}) and (\ref{t2}) can be rewritten in the following 
compact form,
\begin{equation}
\left| \frac{\alpha^2}{e^{2\Lambda}\,r^4} -
\frac{2\,\alpha}{e^{2\Lambda}r^3} -
 \frac{\alpha\, \Lambda'}{e^{2\Lambda}
r^2}  \right|
\leq  \frac{e^{2\omega t}}{(10^{10} {\rm cm})^2}
\end{equation}
\begin{equation}
\left| \gamma^2 \left\{ 
\frac{\alpha}{e^{2\Lambda}\,r^2} + \beta^2\, \left[
 \omega^2 \,e^{2\alpha/r} +
\frac{\Lambda'}{e^{2\Lambda}\,r}\right] +2 \,\beta\,\frac{\alpha \,\omega\,
e^{\alpha/r}}{r\,e^{\Lambda}}\right\}\right| \leq 
\frac{e^{2\omega t}}{(10^{10} {\rm cm})^2}.
\end{equation}
It is seen from these equations that in this expanding ``wormhole
universe'' the tidal forces felt by an observer at fixed $r$ diminish with
time. The traversability
criteria will be then  satisfied for any value of $\omega$
if $t$ is big enough.  On the other hand, when the 
wormhole is contracting ({\em i.e.}
$\Omega (t)=e^{-2\omega t}$), the traversability
criteria constrain the possible values of $\alpha$, $\beta$, and
$\phi_{\rm th}$ more and more as the geometry evolves, 
eventually 
rendering the trip impossible. The behaviour of the tidal
forces is then intimately 
related to the conformal factor $\Omega (t)$, and one
would generically expect that if there is an expansion (contraction), the
tidal 
forces will diminish (grow) with $t$.

\section{Final comments}

We presented the complete analytical solution for a restricted 
class of evolving 
\wh geometries by imposing a suitable form
of the trace of the stress-energy tensor for matter.
This, in turn, depends on the particular election of
the conformal factor. With this solution we have analyzed 
different scenarios of WEC violation, extending in this way
previous works by Kar and Kar and Sahdev \cite{KAR,KAR2}, 
in the sense that the redshift
function in our case is not zero. 
It is worth noting that we have
introduced a nonzero energy flux in the stress-energy tensor, 
which also renders
our non-static solution more general than the ones presented before.

We should also remark that one of the keys for the existence of these
solutions is their non-flat asymptotic behaviour. In an asymptotically   
flat space-time, the existence of traversable \wh solutions in GR is
forbidden by the theorem of topological censorship \cite{fsw}. We have also
studied
some traversability criteria concerning tidal forces that a 
traveler might feel. We found out that,
since the geometry itself is evolving in time, there could be in
some cases an expansion of the throat of the \wh  which may favor
the diminishing of destructive forces upon the trip. 

\acknowledgments

This work has been partially supported by CONICET and UNLP. L.A.A. held
partial support from FOMEC. One of us (SPB) would like to thank the
International Center for Theoretical Physics in Trieste for hospitality.


\newpage
						 
\begin{table}
\caption{Functions involved in WEC inequalities}
\begin{tabular}{c}
$f_1 =\frac 34 \left( \frac{\dot \Omega}{\Omega} \right)^2 $\\
$f_2= -\frac {\ddot \Omega}{\Omega}+
\frac 32 \left( \frac{\dot \Omega}{\Omega} \right)^2 $\\
$f_3= \frac 32 \left( \frac{\dot \Omega}{\Omega} \right)^2  -
\frac {\ddot \Omega}{\Omega}$\\
$g_1= \frac{e^{-2\alpha/r}}{r} \left( 2 e^{-2\Lambda} \Lambda^\prime
+ \frac {1}{r} - \frac{e^{-2\Lambda}}{r} \right) $\\
$g_2=\frac{e^{-2\alpha/r} e^{-2\Lambda} }{r}
\left( 2 \Lambda^\prime
+ \frac{2 \,\alpha }{r^2} \right)$\\
$g_3=\frac{e^{-2\alpha/r}}{r} \left( e^{-2\Lambda} \Lambda^\prime
+ \frac 1r - \frac{ e^{-2\Lambda} }{r}- \frac {\alpha e^{-2\Lambda} }{r^2}+
\frac{ \alpha^2   e^{-2\Lambda} }{r^3} - 
\frac{ \alpha e^{-2\Lambda} \Lambda^\prime}{r} \right)$ \\

\end{tabular}
\label{tabla1}
\end{table}
						
\begin{table}
\caption{Possible conformal functions}
\begin{tabular}{cccccc}
$\Omega(t)$ & $\Upsilon(t)$ &  $f_1$ & $f_2$ & $f_3$ & 
Domain for $t$ \\ \hline
$e^{2\omega t} $&$ -6 \omega^2 $&$ 3 \omega^2 $&$ 2 \omega^2 $&$ 2 \omega^2 $&$ -\infty <t< \infty$\\
$e^{-2\omega t} $&$ -6 \omega^2 $&$ 3 \omega^2 $&$ 2 \omega^2 $&$ 2 \omega^2 $&$ -\infty <t< \infty$\\
$t^{2n} $&$ 6n(1-n) t^{-2}$&$ 3t^{-2} $&$ (2 n^2 + 2n) t^{-2}$&$ (2n^2 +2n) t^{-2} $&$ 0<t<\infty$\\
$ a^2+t^2 $&$ \frac {-6 a^2}{a^2+t^2} $&$ \frac {3 t^2}{(a^2+t^2)^2}$&$
\frac {-2 a^2 + 4  t^2}{(a^2+t^2)^2} $&$ 
\frac {\frac 12 t^2 -2 a^2}{(a^2+t^2)^2} $&$ -\infty < t<\infty $\\
$\sin^2(\omega t) $&$ 6 \omega^2 $&$ 3 \omega^2 \cot^2(\omega t) $&$ 
4 \omega^2 \cot^2 (\omega t) + 2 \omega^2 $&$ 4 \omega^2 \cot^2 (\omega t)+
2 \omega^2 $&$ \frac {m \pi}{\omega}<t<\frac {(m+1) \pi}{\omega}$\\

\end{tabular}
\label{tabla2}
\end{table}


\begin{figure}  
\label{wec}
\begin{center}
    \leavevmode
   \epsfxsize = 10cm
     \epsfysize = 8cm
    \epsffile{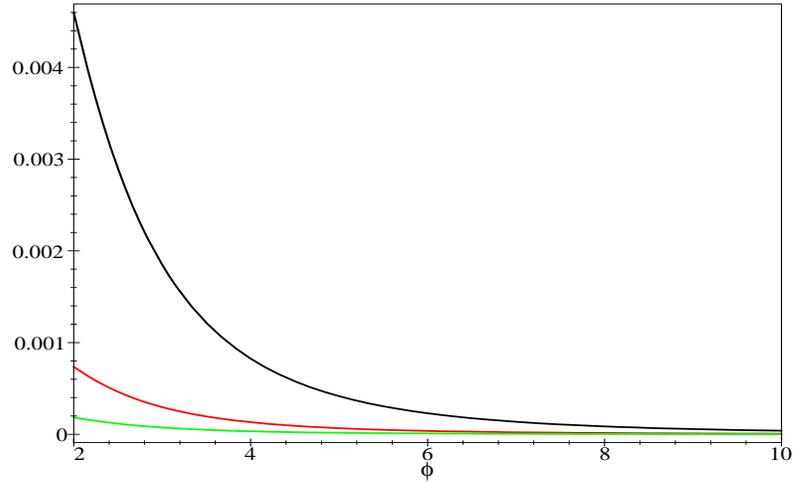}
    \end{center}
\caption{Behavior of the radial part of the first WEC inequality, 
$g_1(\phi)$, near the wormhole throat.  
The curves are, from black to grey, the corresponding
to $\alpha=2,5$ and 10 and all them are presented for the case $\phi \th =2$.}

\end{figure}

\begin{figure}  
\label{wecd}
\begin{center}
    \leavevmode
   \epsfxsize = 10cm
     \epsfysize = 8cm
    \epsffile{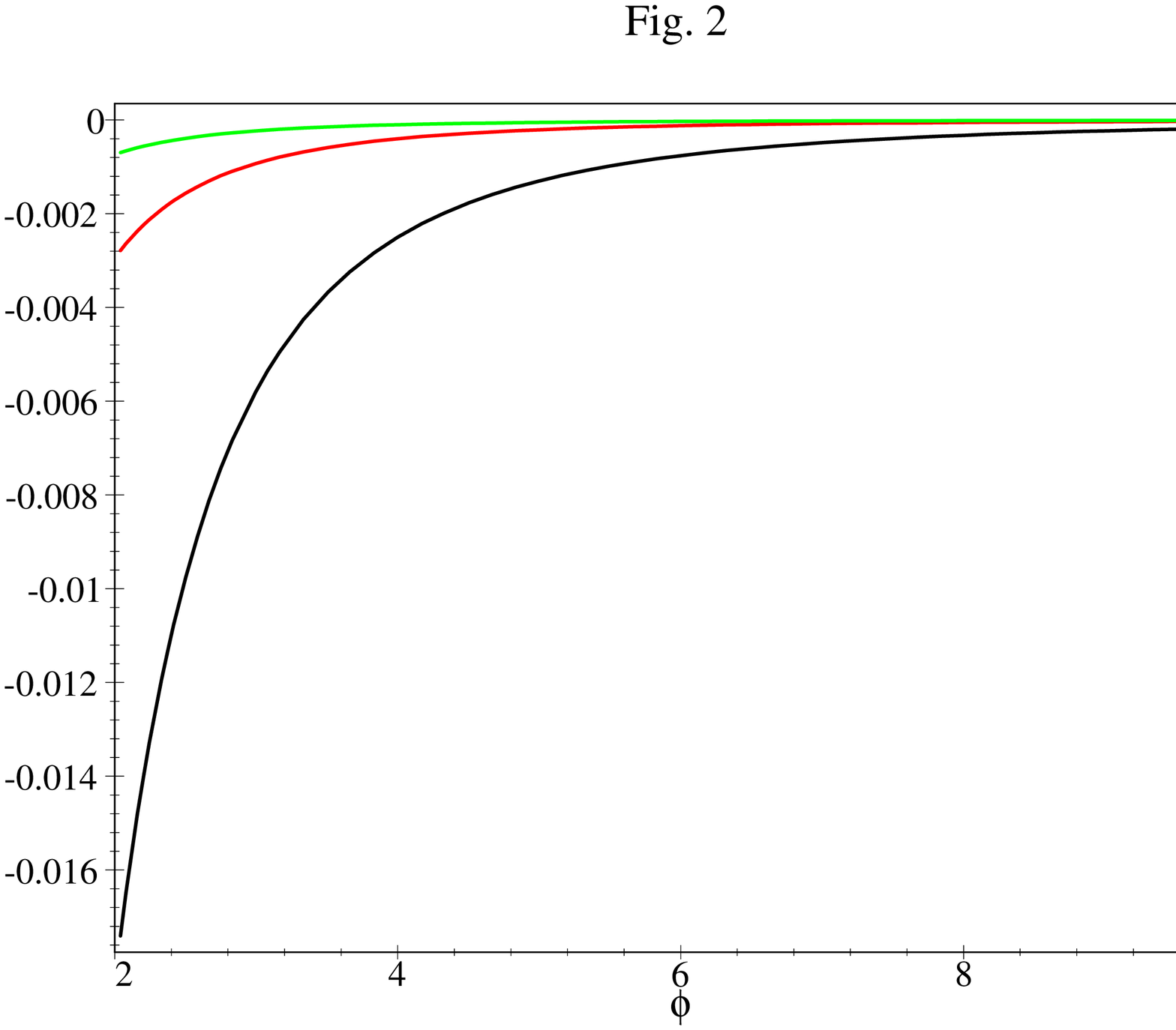}
    \end{center}
\caption{As in Fig. 1 but for the function $g_2 (\phi)$,
involved in the second WEC inequality.}    

\end{figure}

\begin{figure}  
\label{wect}
\begin{center}
    \leavevmode
   \epsfxsize = 10cm
     \epsfysize = 8cm
    \epsffile{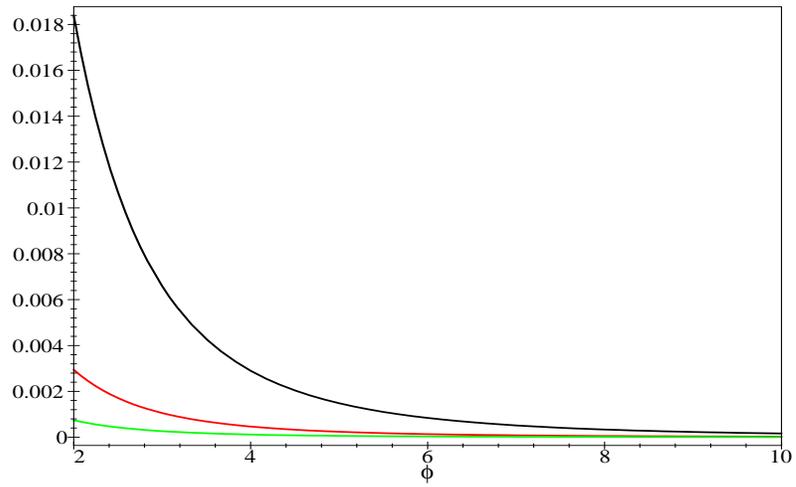}
    \end{center}
\caption{As in Fig. 1 but for the function $g_3 (\phi)$, involved in the
third WEC inequality.}

\end{figure}    
\end{document}